# Freeform metagratings based on complex light scattering dynamics for extreme, high efficiency beam steering

**Authors**: Jianji Yang[1], David Sell[2], and Jonathan A. Fan[1]*

**Affiliations**:

1. Department of Electrical Engineering, Stanford University, Stanford, CA 94305
2. Department of Applied Physics, Stanford University, Stanford, CA 94305

*Address correspondence to: jonfan@stanford.edu

## Abstract

Conventional phased-array metasurfaces utilize resonant nanoparticles or nanowaveguides to specify spatially-dependent amplitude and phase responses to light. In nearly all these implementations, subwavelength-scale elements are stitched together while minimizing coupling between adjacent elements. In this report, we theoretically analyze an alternate method of metasurface design, utilizing freeform inverse design methods, which support significantly enhanced efficiencies compared to conventional designs. Our design process optimizes wavelength-scale elements, which dramatically increases the design space for optical engineering. An in-depth coupled mode analysis of ultra-wide-angle beam deflectors and wavelength splitters shows that the extraordinary performance of our designs originates from the large number of propagating modes supported by the metagrating, in combination with complex multiple scattering dynamics exhibited by these modes. We also apply our coupled mode analysis to conventional nanowaveguide-based metasurfaces to understand and quantify the factors limiting the efficiencies of these devices. We envision that freeform metasurface design methods will open new avenues towards truly high-performance, multi-functional optics by utilizing strongly coupled nanophotonic modes and elements.

**Keywords**: modal analysis, inverse design, metasurfaces, metamaterials, metagrating, coupled nanostructure, multifunction, high efficiency meta-devices

**INTRODUCTION.**

Metasurfaces are wavefront shaping devices that hold great potential in a broad range of applications such as lensing,[1, 2] beam steering,[3-6] polarization control,[7-9] and holography[10-13]. Compared to traditional bulky optical components, metasurfaces have ultra-thin form factors and can readily integrate into ultra-miniaturized optical systems. In addition, they can be engineered to support new types of optical functionality that are difficult to achieve with bulk optics. Examples of such devices include a compact aberration-corrected lens[14] and metasurface-based polarization filters that can significantly improve spatial resolution in single-molecule microscopy.[15]

Initial conceptions of metasurfaces utilized subwavelength-scale dielectric or metal resonators for amplitude and phase control.[5,16-19] While these devices were suitable for demonstrating the metasurface concept, these devices exhibited poor transmission efficiencies at visible and near-infrared wavelengths. An alternative approach to metasurface design has involved the utilization of dielectric nanowaveguides,[1,2,7,20,21] which are stitched together into arrays to produce high-efficiency transmissive devices. High efficiency is possible for these devices because they utilize lossless dielectric materials. Furthermore, with this approach, phase response and transmission efficiency are nearly decoupled parameters. Phase response is obtained by phase accumulation along the waveguide and can be specified through a combination of the waveguide cross-section geometry, relative orientation,[1] and length. Transmission efficiency, on the other hand, is essentially an issue of impedance engineering at the substrate-metagrating and metagrating-air interfaces. As such, there exists a large design space for efficient metasurface engineering.

A fundamental assumption of the metasurface strategies above is that sufficient spacing is required between adjacent elements, whether they are resonators or nanowaveguides, to minimize their optical near-field coupling. These spacing dimensions are typically on the order of $\lambda_0/2$, where $\lambda_0$ is the free space operation wavelength. This requirement sets an upper bound on the packing density of elements within meta-devices. With this stitching approach, state-of-the-art dielectric metalenses with a numerical aperture of 0.8, which corresponds to a maximum collection angle of 53 degrees, have been realized.[1] However, for meta-devices deflecting light to angles larger than 50 degrees, the limited packing density of elements results in a significant drop in efficiency.

Metasurfaces based on inverse design[22] are capable of breaking through the limitations of current metasurface designs based on stitched subwavelength-scale elements. In this Article, we present an analysis of several freeform periodic metasurfaces, i.e. metagratings, which deflect light to very large angles, up to 75 degrees, with theoretical efficiencies near or above 90%. We also present multifunctional designs that can efficiently sort light of two different wavelengths into different diffraction orders, with 80% or greater efficiency for each wavelength. These devices possess curvilinear, freeform layouts and exhibit efficiencies well above the current state-of-the-art.

Our in-depth theoretical analysis of the Bloch modes supported by these metagratings reveals that the light transport in our devices is driven by the intricate scattering dynamics of these Bloch modes and their coupling at the substrate-metagrating and metagrating-air interfaces. We also find quantitatively that our freeform metagratings support a large number of propagating modes that are responsible for the light transport, which contributes to high efficiency operation. In sharp contrast, nanowaveguide-based metasurfaces[1,2,21] support a relatively smaller number of propagating modes, and these modes minimally couple together within the metagrating.

**RESULTS AND DISCUSSION**.

We will analyze dielectric metagratings as a model system for meta-device engineering. Our transmissive optical devices serve as blazed gratings and can deflect incident light at one wavelength to a single diffraction channel, as sketched in Figure 1a. They can also be engineered to deflect different wavelengths to different diffraction channels. To investigate the detailed physical mechanisms of light diffraction, we utilize a coupled Bloch mode analysis technique,[23-25] which analyzes the details of light propagation and scattering dynamics inside the periodic nanostructures and quantifies their impact on optical performance and diffraction efficiency. While brute-force simulations based on either grating solvers (e.g., the Fourier modal method (FMM)[24]) or more general-purpose Maxwell solvers (e.g., FDTD, FEM, etc.) can accurately evaluate diffraction efficiency, they do not elucidate the underlying physics of operation.

In the air and substrate regions above and below the metagrating, the light field can be expanded as plane waves propagating in the directions of the permitted diffraction orders (Figure 1b). Inside the metagrating, the fields naturally expand into a Bloch mode basis, [23-25] due to the periodicity of the grating. The dynamics of the light diffraction process can be described as the following. Incident light propagating through the substrate hits the substrate-metagrating interface of the metagrating, exciting a finite number of propagating Bloch modes and an infinite number of evanescent Bloch modes in the metagrating. The evanescent modes exponentially damp along the vertical direction and, for sufficiently thick metagratings, play a negligible role in light transport. The propagating modes propagate vertically in the metagrating and can bounce between the metagrating-air and substrate-metagrating interfaces (Figure 1b). At each of these interfaces, the modes can undergo a combination of three processes: they can excite other modes through inter-mode coupling, reflect back into the metagrating through intra-mode coupling, or out-couple into free space (Figure 1c). High deflection efficiency occurs when the out-coupled radiation from the modes strongly constructively interfere in the desired diffraction channel. Note that, in Figures 1b and 1c, different colors are used to discriminate different Bloch modes of a metagrating operating at a single wavelength.

These mode scattering processes at the metagrating interfaces can be quantified using coupled Bloch mode analysis. [23-25] We denote the coupling coefficients between the incident plane wave and Bloch modes at the bottom interface as $\mathbf{t}_B$, where the subscript 'B' represents the *bottom* (substrate-metagrating) interface. The coupling coefficients between the Bloch modes and transmitted plane wave in the desired diffraction channel at the metagrating-air interface are $\mathbf{t}_T$, where the subscript 'T' represents the *top* (metagrating-air) interface. If there are N propagating Bloch modes in the metagrating, $\mathbf{t}_B$ and $\mathbf{t}_T$ are represented as N×1 arrays. In addition, two N×N matrices, $\mathbf{S}_B$ and $\mathbf{S}_T$, describe the coupling between Bloch modes as they scatter at the bottom and top metagrating interfaces, respectively. The off-diagonal terms in $\mathbf{S}_B$ and $\mathbf{S}_T$ correspond to inter-mode coupling, while the diagonal terms in the matrices correspond to intra-mode coupling. We note that the Bloch modes are orthogonal and only interact with each other at the metagrating interfaces.

The metagrating can be treated as a generalized Fabry Perot resonator,[23-25] in which the propagating modes experience multiple round trips within the resonator. With each round trip, a fraction of the light from each mode will couple to the desired diffraction channel, as well as many undesired channels. The total field transmitted into the desired diffraction channel, $t$, has contributions from all of the propagating modes and each of their round trips, and is expressed as:

$$t = (\mathbf{t}_T)' \left[ \boldsymbol{\varphi} + \sum_{j=1}^{m} (\boldsymbol{\varphi} \mathbf{S}_B \boldsymbol{\varphi} \mathbf{S}_T)^j \right] \mathbf{t}_B, \text{ (full dynamical model)} \quad (1)$$

The prime next to $\mathbf{t}_T$ denotes a transpose operation. In Equation 1, $\boldsymbol{\varphi}$ is an N×N diagonal matrix with diagonal terms $\varphi_{pp} = \exp(ik_0 n_p h)$, where the subscript '$p$' represents the $p^{th}$ mode. These terms represent the phase accumulated by each propagating mode upon a single pass through the metagrating. $k_0$, $n_p$ and $h$ denote the wavenumber, effective index of the $p^{th}$ mode, and the metagrating thickness, respectively. In Equation 1, the $\mathbf{t}_B$ and $(\mathbf{t}_T)'$ terms multiplied with the diagonal matrix $\boldsymbol{\varphi}$ yields a number that corresponds to the total field transmission due to the single-pass contributions of the propagating modes. The term $\sum_{j=1}^{m} (\boldsymbol{\varphi} \mathbf{S}_B \boldsymbol{\varphi} \mathbf{S}_T)^j$ is a multiple-scattering term that represents the contributions from all the modes after $m$ round trips inside the metagrating.

The single-pass term contains only contributions from propagating modes directly excited by the incident field. The multiple-scattering term, on the other hand, includes both inter-mode coupling and intra-mode coupling contributions, i.e., the light circulating inside the periodic structure. As mentioned above, the intra-mode coupling is represented by the diagonal terms of the matrices $\mathbf{S}_B$ and $\mathbf{S}_T$, while the off-diagonal terms represent the inter-mode coupling. As such, the contributions of single-pass transmission, inter-mode coupling, and intra-mode coupling can be separated and quantified in Equation 1. The approximate transmission due to the single-pass terms is:

$$t \approx (\mathbf{t}_T)' \boldsymbol{\varphi} \mathbf{t}_B. \text{ (single-pass approximation)} \quad (2)$$

The approximate transmission neglecting intermode coupling is:

$$t \approx \left(\mathbf{t}_{T}\right)'\left[\boldsymbol{\varphi} + \sum_{j=1}^{m}\left(\boldsymbol{\varphi}\mathbf{S}_{B}^{\text{diag}}\boldsymbol{\varphi}\mathbf{S}_{T}^{\text{diag}}\right)^{j}\right]\mathbf{t}_{B}, \text{ (multiple-scattering, no intra-mode coupling)} \qquad (3)$$

$\mathbf{S}_{B}^{\text{diag}}$ and $\mathbf{S}_{T}^{\text{diag}}$ are obtained by setting the off-diagonal terms of $\mathbf{S}_B$ and $\mathbf{S}_T$ to zero.

By comparing Equations 1 – 3 for freeform and nanowaveguide-based metasurfaces, we will identify clear distinctions between their underlying physical mechanisms of operation. We emphasize that the expression in Equation 1 accounts for all contributions to beam steering into the desired diffraction channel, and that the single interface coupling coefficients (i.e., $\mathbf{t}_B$, $\mathbf{t}_T$. $\mathbf{S}_B$, and $\mathbf{S}_T$) can be rigorously computed using an open-source FMM package.[26] The major source of error is the numerical error originating from the inevitable truncation of the Fourier harmonic series retained in the FMM computation. For the examples presented in the paper, we have verified that numerical convergence is achieved (see Figure S1 in the Supplementary Section).

**Performance of nanowaveguide-based metagratings**. We begin our theoretical analysis with an examination of titanium dioxide nanowaveguide-based transmissive metagratings,[1,2,7,21] as sketched in Figure 1a. These metagratings are designed to deflect normally-incident unpolarized plane waves with free space wavelength $\lambda_0$=1050nm to the +1 diffraction order channel, and they have thicknesses on the order of $\lambda_0$. They are based on documented methods,[21] and their deflection efficiencies are plotted in Figure 2a (blue dotted line) and are consistent with Ref. 21. According to our recent benchmark study,[22] these designs exhibit slightly higher efficiencies for large deflection angles compared to nanowaveguide-based metagratings designed by other methods.[1,2,7] We note that silicon nanowaveguide-based metagratings [2,7] exhibit similar efficiencies as these titanium dioxide devices (Section 3 of the Supplementary Section).

The plot in Figure 2a shows that these nanowaveguide-based metagratings operate with modest to high efficiencies (70% or greater) at angles less than 50 degrees. However, the efficiencies significantly decrease for larger deflection angles (greater than 50 degrees). The decrease in efficiency at large angles can be qualitatively understood as a result of undersampling due to an insufficient number of nanowaveguides per grating period. At large angles, the periods of these deflectors reduce to length scales just slightly larger than $\lambda_0$. For example, the period of a metagrating supporting a first order diffraction angle at 50 degrees has a period that is ~1.3$\lambda_0$ and can host no more than two nanowaveguides.[1,2,7,21] With only two nanowaveguides, there are insufficient degrees of freedom in the optical design space to engineer both transmission efficiency and phase response.

In the following, we quantitatively analyze the optical properties of a 75-degree metagrating to understand the origins of the low-efficiency performance of large-angle devices. Figure 2b shows that for a TM-polarized (*p*-polarized) incident wave, the deflector supports three propagating modes. The first two modes (M1 and M2) are spatially confined in the wide and narrow pillars, respectively, indicating that the two pillars do not strongly couple in the near-field. The third propagating mode (M3) is mostly distributed in air and has an effective index near 1.0. To understand the role of each mode in the light deflection process, we plot in Figure 2c the deflection efficiency (red circles), predicted by Equation 1 (with $m \rightarrow \infty$), keeping only one mode (M1), two modes (M1 + M2), and all three modes. The efficiency calculated with all three modes agrees well with that calculated rigorously (black dashed line), indicating that all of the modes are essential for device operation. Our analysis for TE-polarized (*s*-polarized) incident light is summarized in Figure S4.

The square of the magnitudes of the scattering parameters ($t_T$, $t_B$, $S_B$ and $S_T$) of the three propagating modes at the top and bottom metagrating interfaces are plotted in Figure 2d. The off-diagonal terms in $|S_B|^2$ and $|S_T|^2$ have small values (less than 0.02, white color), indicating that there is negligible inter-mode coupling at the interfaces. This lack of coupling is consistent with the minimal spatial overlap between the modes, as plotted in Figure 2b. We also find that the values in $|t_B|^2$ are large for all of the modes (red or blue color). In fact, the sum of all the terms in $|t_B|^2$ is greater than 99%, indicating that less than 1% of the incident light gets reflected at the substrate-grating interface. Nanowaveguide-based metagratings operating at other deflection angles and polarizations also generally exhibit small inter-mode coupling and strong impedance matching at the substrate-grating interface (Figure S5).

The transmission of the incident wave into the desired diffraction order, in the single-pass approximation limit, is calculated using Equation 2 and plotted as red circles in Figure 2a. These numbers effectively superimpose with the transmission values calculated rigorously using the FMM solver,[26] indicating that the single-pass limit captures the underlying physics of device operation, and that light does not bounce within the metagrating resonator. This result is consistent with the fact that there is very little reflectivity at the substrate-metagrating interface, due to the lack of intra-mode and inter-mode coupling ($|\mathbf{S}_B|^2$ in Figure 2d is close to zero in all terms).

This analysis of the mode profiles and scattering parameters helps to elucidate the efficiency limitations of large-angle deflection nanowaveguide-based metagratings. First, back-reflection of the propagating modes at the metagrating-air interface (denoted by the red and purple diagonal terms in $|\mathbf{S}_T|^2$) limit the total transmitted power through the 75-degree metagrating to approximately 80% of the incident power. Light back-reflected at the metagrating-air interface does not redirect back to this interface due to the minimal reflectivity at the substrate-metagrating interface.

Second, the large difference between the total transmission and the deflection efficiency indicates that the three Bloch modes fail to strongly constructively interfere at the desired diffraction channel ($+1^{th}$ order) and destructively interfere at the other, non-desired, diffraction channels ($0^{th}$ and $-1^{th}$ order). Rather, there is a substantial amount of light transmitted into the $0^{th}$ order diffraction channel. We theorize that these devices support an insufficient number of propagating Bloch modes, which limits the design space for optical engineering and does not allow the modes to achieve our desired interference conditions. This limitation does not exist for small-angle deflectors, which have larger periods and support more propagating Bloch modes due to the larger number of nanowaveguides per period. As a demonstration, the mode analysis for an 11-degree deflector, which has an efficiency over 80%, is summarized in Figures S6 and S7.

**Freeform metagrating deflectors**. We now examine silicon freeform metagratings based on adjoint-based inverse design.[22] These devices are designed to operate at $\lambda_0$=1050nm and are specified to be 325nm thick (~$0.3\lambda_0$), which is substantially thinner than the nanowaveguide-based metasurfaces above. Our freeform metagrating deflectors display high deflection efficiency for both small and large deflection angles (Figure 3a, blue dotted line); for deflection angles ranging from 10 to 75 degrees, the deflection efficiencies of our designs range from 89% to 95%. To check whether the underlying physics of these metagratings can be described using the single-pass approximation, we calculate the single-pass transmission for these devices using Equation 2 and plot the results in Figure 3a (red circles). These transmission values strongly deviate from those rigorously calculated (blue dots), indicating that our freeform metagratings operate with different physics than the nanowaveguide-based metasurfaces.

We focus on the 75-degree deflector depicted in Figure 3b as a model system for further analysis. The layouts for metagratings operating for other deflection angles are summarized in Figure S8. For TM incident plane waves, the metagrating supports seven propagating modes, plotted in Figure 3b, which is significantly larger than the three modes supported by the nanowaveguide-based device from Figure 2b. Qualitatively, our freeform metagrating supports a sufficient number of propagating Bloch modes to enforce constructive interference at the desired diffraction channel ($+1^{th}$ order) and destructive interfere at the other diffraction channels ($0^{th}$ and $-1^{th}$ order).

The square of the magnitudes of the scattering parameters for this freeform device are plotted in Figure 3c and reveal qualitatively different dynamics than those from the nanowaveguide-based metagratings in Figure 2d. Many of the off-diagonal terms in $|\mathbf{S}_B|^2$ and $|\mathbf{S}_T|^2$ are no longer negligible, indicating the presence of inter-mode coupling at the metagrating interfaces. Inter-mode coupling is likely promoted by the strong spatial overlap between some of the modes (Figure 3b), and mediates new and complex mode dynamics. For example, we find that the incident plane wave does not strongly excite modes M6 and M7 (white color in $|\mathbf{t}_B|^2$). However, due to strong inter-mode coupling, these modes can still be excited within the metagrating and couple to the desired diffraction order channel. In another example, we also find that modes M3 and M6 do not couple efficiently to the desired diffraction order at the metagrating-air interface (white color in $\mathbf{t}_T$). However, these modes can couple with other modes via inter-mode coupling, which then scatter into the desired diffraction channel. We also find that some of the modes (M3 and M7) support strong intra-mode coupling at the top and bottom metagrating interfaces (blue diagonal terms in $|\mathbf{S}_B|^2$ and $|\mathbf{S}_T|^2$) and can experience many round trips within the metagrating.

To quantify how inter-mode coupling, intra-mode coupling, and multiple round trip dynamics contribute to the final metagrating efficiency, we calculate device efficiencies with and without inter-mode coupling as a function of the total number of round trips $m$. The results are summarized in Figure 3d, where we plot the efficiencies using full coupling dynamics calculated from Equation 1 (dotted black line) and dynamics without inter-mode coupling calculated from Equation 3 (solid black line). For $m = 0$, no round trips are accounted for and Equations 1 and 3 both reduce to the single-pass model described by Equation 2. We find that as *m* increases, the prediction by the two equations show dramatically different behavior. The model with full coupling dynamics (dotted black line) shows an oscillatory convergence toward the exact value of ~90% (red line) after approximately 20 round trips. The model that neglects inter-mode dynamics, on the other hand, converges to an incorrect value of only 70%. These data clearly show that inter-mode and intra-mode coupling both play critical roles in our high-efficiency freeform metagratings, and these coupling phenomena mediate pronounced multiple-round-trip dynamics. Our analysis for TE-polarized incident light is summarized in Figure S9 and also displays similar intricate scattering dynamics.

**Multi-Wavelength Function**. There has been tremendous interest in extending metasurface functionality to multiple wavelengths, which would dramatically extend the scope of applications. To date, multi-wavelength lenses[3, 27, 28] and deflectors[3] have been realized by spatially multiplexing[27] or stitching subwavelength-scale elements.[3,28] The specification of wavelength-dependent phase profiles is challenging due to the dispersive nature of nanoscale waveguides and resonators. To date, experimental demonstrations of multi-wavelength devices have yielded only modest efficiencies.

Our optimization strategy can readily generalize to the design of high-efficiency, multiple-wavelength devices. To demonstrate, we design a 325nm-thick silicon metagrating that can efficiently transmit normally incident TE-polarized beams with wavelengths 1μm and 1.3μm to +36 degrees (efficiency ~76%) and -50 degrees (efficiency ~83%), respectively (Figure 4a). We also design and analyze a high-efficiency, polarization-independent wavelength splitter, summarized in Figure S10. To survey the modes of the TE-polarized beam splitter, we plot the mode indices ($n_{\text{eff}}$) of the propagating Bloch modes of the device as a function of wavelength in Figure 4b. At λ=1μm, the metagrating supports seventeen modes, while at λ=1.3μm, the device supports only nine modes, and modes 10-17 are cut off.

An examination of the mode profiles reveals mode-dependent dispersion properties. Modes 1-5 have spatial profiles that exhibit little variation as a function of wavelength. We plot the mode profiles of Mode 1 at the two operating wavelengths in Figure 4c and clearly see that the profiles are nearly the same. Modes 6-17, on the other hand, have spatial profiles that vary strongly as a function of wavelength. Mode 6, for example, possesses an entirely different spatial profile at the two operating wavelengths (Figure 4c). The intricacy of the spatial mode profiles and their dependence on wavelength in high performance meta-devices clearly underscores the need for advanced optimization algorithms in device design.

To characterize the impact of inter-mode and intra-mode coupling on device efficiency, we perform an analysis similar to that in Figure 3d for each operating wavelength of the metagrating. Efficiencies calculated using the full dynamical model (Equation 1, dotted black lines) and the model without inter-mode coupling (Equation 3, solid black lines) are plotted in Figures 4d and 4e as a function of number of round trips. The plots for the full dynamical model show strong oscillatory convergence toward the exact value (red dashed lines), which is consistent with the dynamics of the single-wavelength freeform deflector (Figure 3d). Without inter-mode dynamics, the calculated efficiencies are far from the exact value. These data clearly reveal that strong intra-mode and inter-mode dynamics are central to high efficiency device operation.

**CONCLUSION**

In summary, we have demonstrated that the efficiencies of metasurfaces can be significantly improved using freeform inverse design concepts, compared to the current state-of-the-art. Our analytical models describing light diffraction through the metasurfaces reveal that freeform designs support a large number of propagating Bloch modes, and that these modes undergo multiple scattering at the device interfaces. These scattering processes include inter-mode and intra-mode coupling, and they facilitate multiple-round trip mode dynamics within the device. Metasurfaces based on nanowaveguides, on the other hand, support a smaller number of propagating Bloch modes and can be accurately described using single-pass dynamics.

For meta-devices that steer incident light to a specific angle, high efficiency corresponds physically to strong constructive interference from out-scattered Bloch modes into the desired diffraction channel. In this context, the efficiency enhancements supported in our freeform metasurfaces, compared to nanowaveguide-based devices, can be understood as follows. First, the multiple scattering processes supported in these devices yield a wider design space for specifying the amplitude and phase response of each mode. Second, these devices support a larger overall number of modes, which provides more degrees of freedom for enforcing constructive interference into the desired diffraction channel and for impedance engineering at the metagrating interfaces, pending that the modes are properly optimized. We envision that metasurfaces based on freeform designs will serve as a platform for studying the limits of subwavelength-scale mode engineering in high-efficiency, multi-functional diffractive optical systems.

**Methods.**

**Adjoint-based topology optimization procedure**. The details of our design methodology are documented in Ref 22 and other related sources.[29] We define a Figure of Merit (FoM) corresponding to the specified device function. During each iteration, the electromagnetic fields within the device are utilized to evaluate the gradient of the FoM with respect to the dielectric constant at every spatial location within the device. The information of gradient of FoM drives the FoM toward the target. Our method can readily extend to multiple functionalities (e.g., polarizations, wavelengths) and parallelize the evaluation of gradients of the FoM's of each functionality.[22]

**Robustness of the freeform metagratings to fabrication imperfections**. We incorporate robustness control[22, 29] into our optimization algorithms. The robustness control algorithms ensure that our designs are not sensitive to "over-etching" and "under-etching" imperfections in fabricated devices. For our 75-degree metagrating deflector in Figure 3, we examine the effects of "over-etching" and "under-etching" in Figure S11. Owing to the robustness control, the over-etched and under-etched patterns both show very high efficiency (>83%). Fabricated device designs that incorporate robustness control have performance metrics close to their theoretical predicted values. For example, experimental characterization of a fabricated wavelength splitter (design in Figure 4) shows that the measured device efficiency agrees well with the theoretical prediction.[22]

**Acknowledgements.** The simulations were performed in the Sherlock computing cluster at Stanford University. This work was supported by the U.S. Air Force under Award Number FA9550-15-1-0161, the Office of Naval Research under Award Number N00014-16-1-2630, and the Alfred P. Sloan Foundation. D. Sell was supported by the National Science Foundation (NSF) through the NSF Graduate Research Fellowship. The authors thank Sage Doshay for a critical reading of the manuscript.

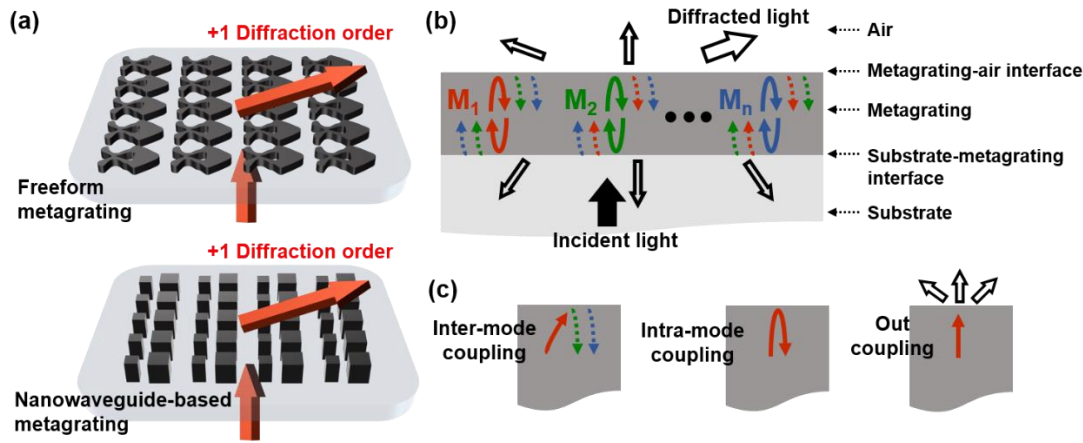

**Figure 1**. **Scattering dynamics of light in a metagrating**. (**a**) Sketches of a freeform metagrating (top) and a nanowaveguide-based metagrating (bottom) that deflect normally-incident monochromatic light to the +1 diffraction order. (**b**) Schematic of scattering dynamics within a metagrating, for normally-incident monochromatic light. Inside the metagrating, energy is carried by multiple propagating Bloch modes, $M_i$, which bounce between the top (metagrating-air) interface and the bottom (substrate-metagrating) interface. (**c**) Schematic of scattering dynamics at the metagrating-air interface. When the propagating modes hit the interface, they can excite other modes (inter-mode coupling, dotted curved lines, left), be back-reflected (intra-mode coupling, solid curved line, middle), or couple into free-space diffraction channels (out-coupling, hollow arrows, right). Similar scattering processes exist at the substrate-metagrating interface. In (b) and (c), all the modes have the same wavelength/frequency, and the different colors are used to differentiate different Bloch modes.

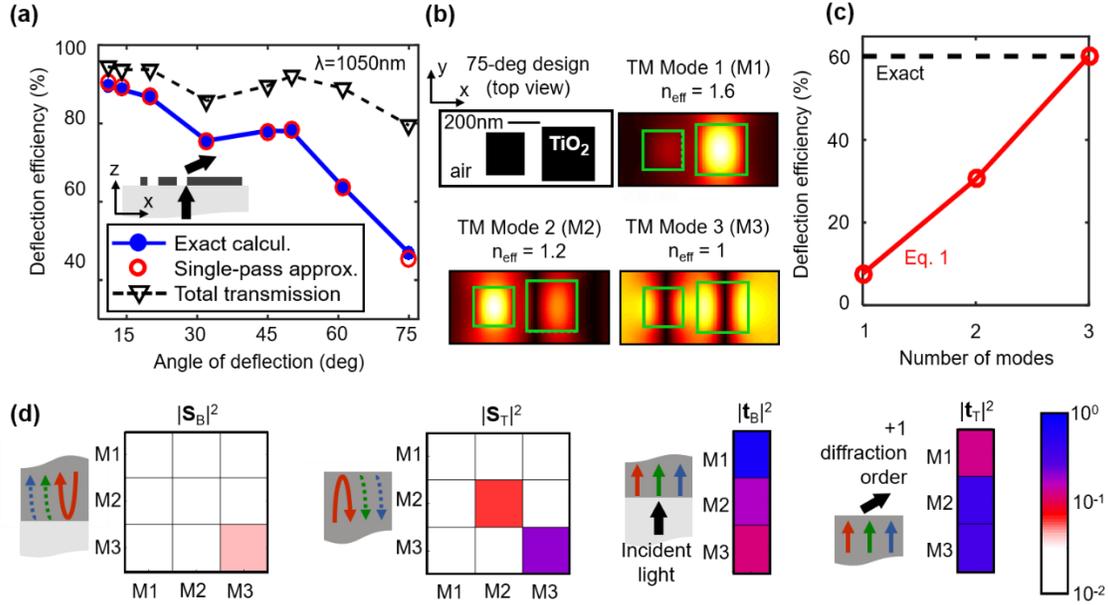

**Figure 2. Performance of nanowaveguide metagrating deflectors for different deflection angles**. (**a**) Deflection efficiency (blue line) of titanium dioxide nanowaveguide-based metagratings, illuminated by normally incident unpolarized light, as a function of deflection angle, obtained by FMM calculation.[24] The red circles represent the transmission obtained using the single-pass approximation in Equation 2. The black triangles represent the total transmission into all the transmitted diffraction orders, not just the desired order. (**b**) Mode analysis of a metagrating that deflects TM-polarized (*p*-polarized) incident light to a 75 degree angle. The intensity plots are of the mode profiles ($|H_y|^2$) of all three propagating Bloch modes. The green lines outline the transverse cross-sections of nanowaveguides. (**c**) Deflection efficiencies of the device from (b), calculated using Equation 1, for 1, 2, and 3 propagating modes retained in the model. The black dashed line represents the efficiency obtained by rigorous FMM calculation. (**d**) Values of $|t_B|^2$, $|t_T|^2$, $|S_B|^2$, and $|S_T|^2$. The coloration scheme is based on the log scale shown at the far right.

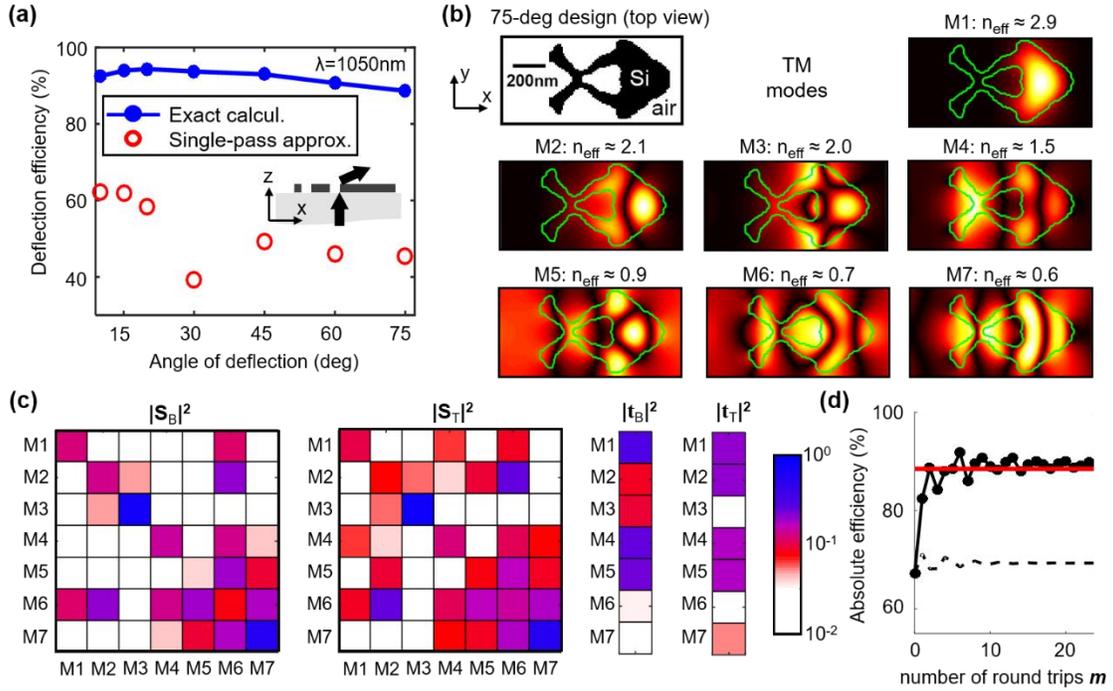

**Figure 3. Performance of freeform metagrating deflectors for different deflection angles**. (**a**) Deflection efficiency (blue dotted line) of freeform silicon metagratings as a function of deflection angle, illuminated by normally-incident unpolarized light. The red circles represent efficiencies calculated using the single-pass approximation. (**b**) Propagating Bloch mode profiles of a 75-degree metagrating for a TM-polarized incident wave. Green lines outline the silicon nanostructure. (**c**) Values of $|\mathbf{t}_B|^2$, $|\mathbf{t}_T|^2$, $|\mathbf{S}_B|^2$, and $|\mathbf{S}_T|^2$ of the metagrating from (b). (**d**) Deflection efficiency as a function of number of round trips *m*. The solid black line represents efficiencies calculated using the full dynamical model (Equation 1), while the dashed black line represents calculations that neglect inter-mode coupling (Equation 3). The red line represents the fully rigorous value for device efficiency. See Figure S9, the mode analysis of the same metagrating for TE polarization.

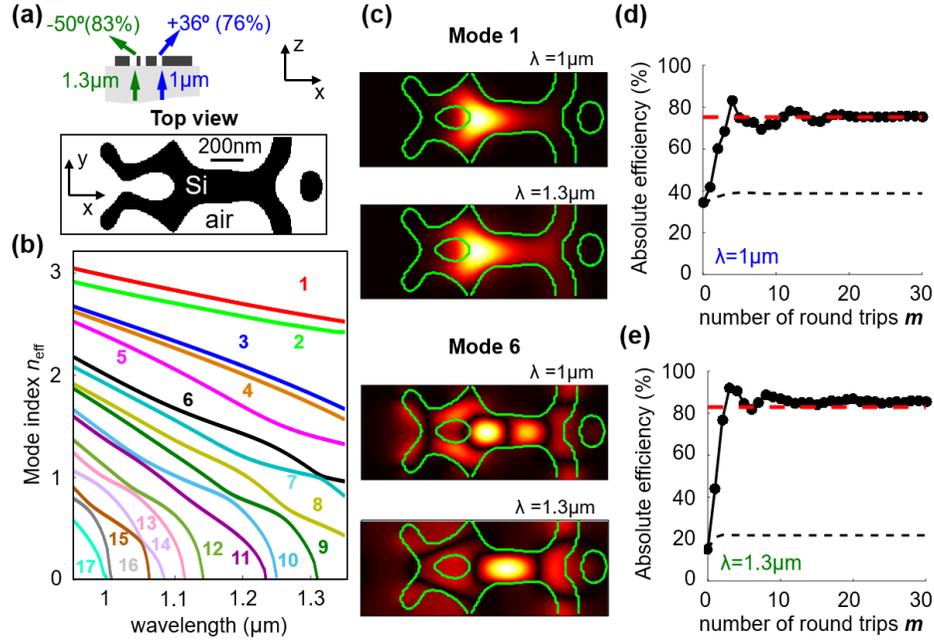

**Figure 4. Analysis of a freeform wavelength-splitting metagrating.** (**a**) Schematic of the freeform metagrating wavelength splitter, which sorts λ=1μm and λ=1.3μm incident light into the +1 and -1 diffraction orders, respectively. (**b**) Effective indices ($n_{\text{eff}}$) of the seventeen propagating modes supported by the metagrating within the wavelengths of interest. At λ=1μm, the device supports Modes 1-17, and at λ=1.3μm, the device supports Modes 1-9. (**c**) Mode profiles ($|\mathbf{H}_y|^2$) of Mode 1 and Mode 6 at the two operation wavelengths. (**d, e**) Deflection efficiency as a function of number of round trips *m*, for (**d**) λ=1μm and (**e**) λ=1.3μm. The solid black lines represent efficiencies calculated using the full dynamical model (Equation 1), while the dashed black lines represent calculations that neglect inter-mode coupling (Equation 3). The red lines represent the fully rigorous values for device efficiency.

19